\documentclass[twocolumn,showpacs,preprintnumbers,amsmath,amssymb]{revtex4}

\usepackage{graphicx}
\usepackage{dcolumn}
\usepackage{bm}

\begin{document}


\title{Numerical and Monte Carlo Bethe ansatz method: 1D Heisenberg model}

\author{Shi-Jian Gu}
\altaffiliation{Email: sjgu@zimp.zju.edu.cn\\URL:
http://zimp.zju.edu.cn/$\widetilde{\;\,}$sjgu/}
\affiliation{Zhejiang Institute of Modern Physics, Zhejiang
University, Hangzhou 310027, P.R. China}

\author{N. M. R. Peres}

\affiliation{Departamento de F\'{\i}sica
e Centro de
F\'{\i}sica da Universidade do Minho, Campus Gualtar, P-4700-320
Braga, Portugal.}

\author{You-Quan Li}
\affiliation{Zhejiang Institute of Modern Physics, Zhejiang
University, Hangzhou 310027, P.R.China}

\date{\today}

\begin{abstract}
In this paper we present two new numerical methods for studying
thermodynamic quantities of integrable models.
As an example  of the effectiveness of these two
approaches, results from numerical solutions of all sets of Bethe
ansatz equations, for  small Heisenberg chains, and Monte Carlo
simulations in quasi-momentum space, for a relatively larger chains, are
presented. Our results  agree with those
obtained by thermodynamics Bethe ansatz (TBA) and Quantum
Transfer Matrix (QTM).
\end{abstract}

\maketitle

\section{Introduction}
\label{mcba_sec:intr}

The study of exactly solvable models is a very important field in
condensed matter physics, which began with  Bethe's solution of
the isotropic Heisenberg chain \cite{HABethe31}. In general, the
Bethe ansatz (BA) solution of a model has several drawbacks: it
has a complex mathematical structure; the excitations are not
immediately available; and most important, it does not give
explicit results even for the thermodynamic quantities of the
system. It was only when Yang and Yang \cite{CNYang69} presented a
strategy to study the thermodynamics of BA solvable systems that
the temperature dependence of quantities such as the specific heat and
the magnetic susceptibility become available. The method is now
designated as thermodynamic Bethe ansatz  and has known many
developments in the last thirty years \cite{MTakahashib}.
Additionally, correlation functions, such as the conductivity, can
not be obtained from the BA equations alone, and  combination
of BA results
with other methods is required for their calculation \cite{prl}.

The BA ansatz method has been applied to Bose, Fermi
 \cite{EHLieb63,CNYang67,EHLieb68}, and spin
systems \cite{HABethe31,ROrbach58}. It is a general feature of the
BA solution, first proved by Yang and Yang\cite{CNYang69} for the
Bose case, that a given eigenstate of the model is characterized
by a unique set of quantum numbers $\{I_j\}$. Further, it also can be
shown that all configurations of these quantum numbers $I_j $
exhaust the Hilbert space of a given model. Since the energy eigenvalues are
functions of the above quantum numbers, instead of using TBA  and
quantum Monte Carlo approaches, we can study BA solvable models
in quantum
number space by classical Monte Carlo. Furthermore, for a small
system (but for larger systems than those available
to exact diagonalization methods),
it is possible to solve the BA equations for
all eigenvalues. Therefore, the
expectation value of an Hermitian operator in thermal equilibrium
can be computed.

In this paper, we shall introduce two numerical approaches for
computing thermodynamic quantities of
Bethe
ansatz solvable models. The methods are ilustrated
with the 1D isotropic Heisenberg model, since this model is well
studied in the literature. Furthermore, the study of the
Heisenberg model is it self relevant, since this system
predicts many properties
of quasi-one-dimensional materials
\cite{NMotoyama96,STakagi96,KRThurber01}. This model
has been investigated by many kinds of methods. For example, the
low temperature behaviors are quite well understood by a
combination of Bethe ansatz \cite{RJBaxterb} and conformal field
theory \cite{AABelavin84,JLCardyb}. A strong logarithm singularity
in the susceptibility at low temperature was first found by the
Bethe ansatz calculation of quantum transfer
matrix \cite{SEggert94} and then verified
experimentally \cite{NMotoyama96,STakagi96}. The thermodynamics of the
model has been studied by
TBA \cite{MTakahashib,MTakahashi71,MGaudin71,MTakahashi85,PSchlottmann85,MTakahashibb,MShiroishi02}
as well as by QTM \cite{MSuzuki87,TKoma89,AKlumper93,AKlumper98}.

The paper is
organized as follows. In Sec. \ref{mcba_sec:iso}, we first briefly
review the BA solution of the isotropic Heisenberg model. In Sec.
\ref{mcba_sec:nba} and Sec. \ref{mcba_sec:mcba}, we introduce the
basic idea of numerical Bethe ansatz (MBA) and Monte Carlo Bethe ansatz
(MCBA).
In Sec.\ref{mcba_sec:spce}
we check the effectiveness of these
two methods  computing  the
specific heat and the magnetic susceptibility in the absence of an
external magnetic
field and compare our results with those obtained from the TBA.
We than use our methods to study the two quantities above in the presence
of an external magnetic field.
A
brief summary is given in Sec. \ref{mcba_sec:summary}.

\section{Isotropic Heisenberg Model}
\label{mcba_sec:iso}

Now let us first review the Bethe ansatz solution of the 1D Heisenberg
chain, which can be found in the book of
Takahashi \cite{MTakahashib}. The Hamiltonian of the
isotropic Heisenberg model is
\begin{eqnarray}
{\cal H}=-J\sum_{l=1}^N\left( S_l^x S_{l+1}^x + S_l^y S_{l+1}^y
+S_l^z S_{l+1}^z \right), \label{mcba_eq:Hamitonian}
\end{eqnarray}
where $N$ is the number of sites, $S_l^x, S_l^y,S_l^z$ are spin
$1/2$ operators at site $l$ and $J=-1, 1$ represent
anti-ferromagnetic and ferromagnetic cases, respectively. The
solution with periodic boundary condition
$\vec{S}_{N+1}=\vec{S}_1$ using the string hypothesis takes the form
\begin{eqnarray}
N\theta(x_\gamma^n/n)=2\pi
I_\gamma^{(n)}+\sum_{m,\beta\neq n,\gamma}
\Theta_{nm}(x_\gamma^n - x_\beta^m). \label{mcba_eq:BAE}
\end{eqnarray}
Here $\theta(x)=2\tan^{-1}(x)$, and

\begin{widetext}
\begin{eqnarray}
\Theta_{nm}(x)&=&\theta\left(\frac{x}{|n-m|}\right)+2
\theta\left(\frac{x}{|n-m|+2}\right) + \cdots +2
\theta\left(\frac{x}{n+m-2}\right)+\theta\left(\frac{x}{n+m}\right)
\;\;\;{\rm For:} \; n\neq m \nonumber \\
&=& 2\theta\left(\frac{x}{2}\right)+2
\theta\left(\frac{x}{4}\right)+\cdots + 2
\theta\left(\frac{x}{2n-2}\right) +\theta\left(\frac{x}{2n}\right)
\;\;\; {\rm For: }\; n=m
\end{eqnarray}
\end{widetext}
and $x_\gamma^n$ is the real part of the n-string which is designated
by
$$
x_\gamma^{n, j}=x_\gamma^n + i(n+1-2j),\;\;j=1, \dots, n
$$
$I_\gamma^n$ is the quantum number of $\gamma$th n-string
(note that $n$ and $\gamma$ are indices). We
denote the number of the n-string by $\alpha_n$, thus $\gamma=1,\dots,
\alpha_n$ and the string configuration $\{\alpha\}$ satisfy
\begin{eqnarray}
\alpha_1 +2 \alpha_2+ \cdots + (M-1)\alpha_{M-1} + M\alpha_M=M,
\label{mcba_eq:stringcondition}
\end{eqnarray}
where $M$ is the number of down spins.
The quantum number of n-string $I_\gamma^{n}$ is an integer
(half-odd integer) if $N-\alpha_n$ is odd (even) and satisfy
\begin{eqnarray}
|I_\alpha^n|\leq \bigl( N-1-\sum_{m=1}^M t_{nm}\alpha_m\bigr)/2,
\label{mcba_eq:QNcondition}
\end{eqnarray}
where $t_{nm}\equiv 2\min (n, m)-\delta_{nm}$. For a given set of
$\{I_\gamma^n\}$, Eq. (\ref{mcba_eq:BAE}) can be solved
numerically and the energy is given by
\begin{eqnarray}
E\{I_\gamma^n\}=-NJ/4+\sum_{n,
\gamma}\frac{2Jn}{(x_\gamma^n)^2+n^2}, \label{mcba_eq:BAenergy}
\label{energ}
\end{eqnarray}
which represents the energy of the lowest weight state in SU(2)
irreducible space designated by $S=N/2-M, S_z=S, S-1, \dots, -S$.
In the presence of an external field $h$ a
Zeeman term is added to Eq. (\ref{energ}).
Hence
the total energy of a given quantum number configuration is given
by
\begin{eqnarray}
E=E\{I_\gamma^{(n)}\}-h\mathcal{M}\,, \label{mcba_eq:tBAenergy}
\end{eqnarray}
where $\mathcal{M}=2S_z$ is the magnetization of the state
\section{Numerical Bethe Ansatz}  %
\label{mcba_sec:nba}              %

In statistical mechanics, the expectation value of an Hermitian
operator $Q$ in thermal equilibrium is given by
\begin{eqnarray}
\langle Q\rangle=\frac{1}{Z} \sum_{\mu}Q_\mu e^{-\beta E_\mu}.
\label{mcba_eq:expectationQ}
\end{eqnarray}
where $Z$ is known as partition function, defined as
\begin{eqnarray}
Z=\sum_{\mu}e^{-\beta E_\mu}\,,
\label{mcba_eq:partitionfunction}
\end{eqnarray}
$\beta$ is inverse
temperature, and $\sum_\mu$ represents sum over all possible
eigenstates of the Hamiltonian.
It turns out that the  variation of $Z$ with respect to
temperature or any other external parameters affecting the system
can virtually tell us everything we might want to know about the
macroscopic behavior of the system. For example, the internal
energy is given by
\begin{eqnarray}
U=\frac{1}{Z}\sum_{\mu}E_\mu e^{-\beta E_\mu}
\end{eqnarray}
From Eq. (\ref{mcba_eq:partitionfunction}), it is easy to see that
the internal energy can also be written in terms of a derivative
of the partition function:
\begin{eqnarray}
U=-\frac{1}{Z}\frac{\partial Z}{\partial \beta}=-\frac{\partial
\ln Z }{\partial \beta}.
\end{eqnarray}
The specific heat is given by the derivative of the internal
energy:
\begin{eqnarray}
C_v=\frac{\partial U}{\partial T}=-k_B \beta^2\frac{\partial
U}{\partial \beta}=-k_B\beta^2\frac{\partial^2 \ln Z}{\partial
\beta^2}.
\end{eqnarray}
where $k_B$ is the Boltzmann constant which is set to unit
hereafter.

Our aim is to combine the idea of statistical mechanics mentioned
above with the numerical solution of the BA equations.
The main idea of the numerical
Bethe ansatz method we introduce here
is, first, to compute all eigenvalues of a BA solvable
model from its corresponding BA equations. Then to compute the
expectation value of the Hermitian operators, representing the
physical observables we are interested in,
by averaging those operators over all states of the system,
weighting each state with its own Boltzmann
weight.

It has been shown \cite{MTakahashib} that the Hilbert space of the
isotropic Heisenberg model is complete under the string
classification. Here we want to show how to travel through all
$C^N_{N/2}$ states in quantum number space and illustrate it by
considering a system of 6 sites.

For the case of $M$ down spins, the first task is how to obtain
all string configurations fulfilling the restriction
(\ref{mcba_eq:stringcondition}). We adopt a time-like number
``$\alpha_M: \alpha_{M-1}: \dots :\alpha_2:\alpha_1$", where the
magnitude $\alpha_n$ measures from 0 to $[M/n]$ (here $[x]$
returns the truncated integer value of $x$), just like hour and
minute in ``HH:MM" measure from 0 to 23 and 0 to 59 respectively.
If we increase the number ``$\alpha_M: \dots :\alpha_1$" by adding
1 to the first digit $\alpha_1$ step by step, we can travel
through all possible values. Among all these numerical values,
only those whose digits satisfy the condition
(\ref{mcba_eq:stringcondition}) are what we need. Then all string
configurations can be found by this procedure. Of course, these
operations  are
realized in a  computer.
In order to make the method clear, let us consider a problem of 6
sites.

{\bf M=0:} It is easy to have the state with all spins up, i. e.
$M=0$, which has energy $E=-JN/4$.

{\bf M=1:} In this case, we only have one string configuration
$\alpha_1=1$ and one quantum number $-2\leq I_1 \leq 2$, thus
there are 5 states. Each of them is represented by one quantum
number in the interval
$[-2, 2]$. We can get all possible quantum number
configurations from the following figure,
\begin{eqnarray}
--\circ--\circ--\bullet--\circ--\circ-- \nonumber
\end{eqnarray}
where the dot is the occupied quantum number, and the open circles
represent other possible quantum numbers. Then the BA equation is
just
\begin{eqnarray}
6\tan^{-1}x^1_1=\pi I_1,
\end{eqnarray}
which has a simple solution $x_1=\tan(\pi I_1/6)$.

{\bf M=2:} Here the string configuration is characterized by
$\{\alpha_1, \alpha_2\}$. We construct a number ``$\alpha_2
:\alpha_1$", in which the maximum value of $\alpha_1$ is 2 ($[2/1]=2$), and
$\alpha_2$ 1 ($[2/2]=1$). Increasing $\alpha_1$ step by step  we generate all
possible configurations of the
 ``$\alpha_2:\alpha_1$" number,
ranging from 0:0 to 1:2. Among  all the generated configurations,
we are only interested in those
satisfying the condition $\alpha_1+2\alpha_2=2$. The
first case is $\alpha_1=2, \alpha_2=0$, in which the quantum
number satisfy $-1.5\leq I_1^1, I_2^1 \leq 1.5$, the second one is
$\alpha_1=0, \alpha_2=1$, in which the number satisfy $-1 \leq
I^2_1 \leq 1$. They can be characterized by
\begin{eqnarray}
--\circ--\bullet--\bullet--\circ-- \nonumber
\end{eqnarray}
and
\begin{eqnarray}
---\circ--\ddagger--\circ--- \nonumber
\end{eqnarray}
respectively, where $\ddagger$ denotes the occupation for a
quantum number of 2-string. In Table \ref{mcba_tab:M2QNF}, we
list all quantum number configurations for $M=2$. The BA equations
for these two cases are
\begin{eqnarray}
6\tan^{-1}x^1_1=&&\pi I^1_1 +\tan^{-1}\frac{x^1_1-x^1_2}{2},\nonumber \\
6\tan^{-1}x^1_2=&&\pi I^1_2 +\tan^{-1}\frac{x^1_2-x^1_1}{2},
\end{eqnarray}
and
\begin{eqnarray}
6\tan^{-1}(x^2_1/2)=\pi I^2_1,
\end{eqnarray}
respectively.

\begin{table}
 \caption{All quantum number configurations for
$M=2$.} \label{mcba_tab:M2QNF}
\begin{center}
\begin{ruledtabular}
\begin{tabular}{|c|c|c|c|c|c|c|c|}
$\alpha_1=2$,
$\alpha_2=0$&$\,I_1^1\,$&$\,$-1.5$\,$&$\,$-1.5$\,$&$\,$-1.5$\,$&$\,$
-0.5$\,$&$\,$-0.5$\,$&$\,$0.5 \\
 & $\,I_2^1\,$ &$\,$-0.5$\,$&$\,\,$0.5$\,$& $\,\,$1.5$\,$
 & $\,\,$0.5$\,$ & $\,$1.5$\,$ & 1.5 \\
\hline $\alpha_1=0$, $\alpha_2=1$ & $I_1^2$ & -1 & 0 & 1 & - & - &
- \\
\end{tabular}
\end{ruledtabular}
\end{center}
\end{table}

{\bf M=3:} In this case the string configuration is characterized
by $\{\alpha_1, \alpha_2, \alpha_3\}$. In the same way as we did
above, we construct a number ``$\alpha_3:\alpha_2:\alpha_1$", the
maximum value for each digit from left to right is 1, 1, 3
respectively. Then we have 3 string configurations with the
condition $\alpha_1+2\alpha_2 +3\alpha_3=3$, which correspond to
the following sequences
\begin{eqnarray}
a:\;\;\; ----\bullet--\bullet--\bullet---- \nonumber
\end{eqnarray}

\begin{eqnarray}
b:&&\;----\circ--\bullet--\circ---- \nonumber \\
&&-------\ddagger------- \nonumber
\end{eqnarray}

\begin{eqnarray}
c:\;\;\; -------\S------- \nonumber
\end{eqnarray}
where $a, b, c$ have 1, 3, 1 states respectively, $\S$ denotes the
site for 3-string. And in Table \ref{mcba_tab:M3QNF}, we list all
quantum number configurations for $M=3$, whose BA equations are
\begin{eqnarray}
6\tan^{-1}x^1_1&=&\pi I^1_1+\tan^{-1}(x_1^1-x^1_2) \nonumber \\
&&+\tan^{-1}(x_1^1-x^1_3),\nonumber \\
6\tan^{-1}x^1_2&=&\pi I^1_2+\tan^{-1}(x_2^1-x^1_1) \nonumber \\ &&
+\tan^{-1}(x_2^1-x^1_3),\nonumber \\
6\tan^{-1}x^1_3 &=& \pi I^1_3+\tan^{-1}(x_3^1-x^1_1) \nonumber \\
&& +\tan^{-1}(x_3^1-x^1_1).
\end{eqnarray}

\begin{eqnarray}
6\tan^{-1}x^1_1 &=& \pi I^1_1+\tan^{-1}(x^1_1-x^2_1) \nonumber
\\ &&
+\tan^{-1}((x^1_1-x^2_1)/3),\nonumber
\\
6\tan^{-1}x^2_1 &=& \pi I^2_1+\tan^{-1}(x^2_1-x^1_1) \nonumber
\\ &&
+\tan^{-1}((x^2_1-x^1_1)/3),\nonumber
\\
\end{eqnarray}
and
\begin{eqnarray}
6\tan^{-1}x^3_1=\pi I^3_1.
\end{eqnarray}
respectively.

\begin{center}
\begin{table}
\caption{All quantum number configurations for $M=3$}
\begin{center}
\vspace{3mm}
\begin{ruledtabular}
\begin{tabular}{|c|c|c|c|c|}

 $\alpha_1=3$,
$\alpha_2=0$, $\alpha_3=0$
 & $\,I_1^1\,$ & -1.0 & - & - \\
 & $\,I_2^1\,$ &   0  & - & - \\
 & $\,I_3^1\,$ &  1.0 & - & - \\
\hline $\alpha_1=0$, $\alpha_2=0$, $\alpha_3=1$
 & $\,I_1^1\,$ & $\,$-1.0$\,$& 0 & 1.0 \\
 & $\,I_1^2\,$ & $\,\,$0$\,$ & 0 & 0 \\
\hline $\alpha_1=1$, $\alpha_2=1$, $\alpha_3=0$
 & $\,I_1^3\,$ & $\,$ 0$\,$& - & - \\
\end{tabular}
\end{ruledtabular}
\end{center} \label{mcba_tab:M3QNF}
\end{table}
\end{center}
\vspace{1mm}

As a result we have totally $C^6_3=20$ distinct configurations of
quantum number whose Hilbert space is complete.

Then, we compute the eigenvalue for a given quantum number
configuration $\{I_\gamma^n\}$ by solving the BA equations
numericaly. For the Heisenberg chain, the BA equations can be solved by
iteration, for other models, such as the  Hubbard model, the BA
equations can be solved by a gradient method.

\section{Monte Carlo Bethe Ansatz}
\label{mcba_sec:mcba}

For a system of $N$ sites, there are $C^N_{N/2}$ quantum number
configurations. This number increases exponentially with the size
of the system, so it is impossible to calculate all eigenvalues
for a large system, such as $N>40$, under present computer
capacity.
This restriction can be overcome by a Monte Carlo method. There are
many Monte Carlo methods available, and we introduce below a new
method that we call Monte Carlo Bethe ansatz. This method
is  a classical Monte
Carlo strategy applied to a quantum problem. The basic idea behind
the MCBA method  is to simulate the random thermal fluctuation of
the system from state to state in quantum number space of the BA
solution. This method is not limited by the sign problem, that
may show up in the usual quantum Monte Carlo methods.

Since the energy eigenvalues are a function of both $M$ and of the
quantum numbers $I_\gamma^n$ we can follow a classical Monte Carlo
strategy, by sampling the configuration space of $M$ and
$\{I_\gamma^n\}$. We now explain how to implement the Monte Carlo
calculation, which follows three steps. Let us assume the present
state is $\mu$ with a corresponding $M_\mu$ --
the number of down spins in state $\mu$. From the state $\mu$
any other state $\nu$ with $M_\nu$, in number of $C^N_{N/2}-1$,
can be obtained.

{\bf step one:} first we choose $M_\nu$, knowing that  the number
of states with $M_\nu$ spins down is $C^N_{M_\nu}-C^N_{M_\nu-1}$,
thus the probability of selecting $M_\nu$ is
$(C^N_{M_\nu}-C^N_{M_\nu-1})/C^N_{N/2}$.

{\bf step two:} selected $M_\nu$, all possible string
configurations given $M_\nu$ are determined from of Eq.
(\ref{mcba_eq:stringcondition}) which satisfy \cite{MTakahashib}
\begin{eqnarray}
\sum_{\alpha_1 + \cdots + M\alpha_M=M}
D(\{\alpha_n\})=(C^N_M-C^N_{M-1})\,,
\end{eqnarray}
with $D(\{\alpha_n\})$ is the number of states, characterized by
the set of quantum numbers $\{I^n_{\alpha}\}$ associated with the
string configuration $\{\alpha_n\}$, and reads
\begin{eqnarray}
D(\{\alpha_n\})=\prod_{i=1}^M C^{N-\sum_{j=1}^M
t_{ij}\alpha_j}_{\alpha_j}\,.
\end{eqnarray}
So, in step two, we select a string configuration with the
probability $D(\{\alpha_n\})/(C^N_M-C^N_{M-1})$.

{\bf step three:} having determined the string configuration, we
then select at random a quantum number configuration, which is the
state $\nu$ we want,  for the given string configuration. From the
partition function $Z$, the  probability density for a  state $\mu$ is
\begin{eqnarray}
p_\mu=(N-2M_\mu+1)e^{-\beta E_\mu},
\end{eqnarray}
where the degenerancy of state $\mu$ was taken into account.
The detailed balance condition  tells us the transition
probability should satisfy
\begin{eqnarray}
\frac{p_\nu}{p_\mu}=\frac{(N-2M_\nu+1)}{(N-2M_\mu+1)}e^{-\beta(E_\nu-E_\mu)}.
\end{eqnarray}
Hence it is possible to use the Metropolis algorithm for the
acceptance ratio to accept or reject the state $\mu$ according to
\begin{eqnarray}
A(\mu\rightarrow\nu)=\left\{\begin{array}{ll}
\frac{(N-2M_\nu+1)}{(N-2M_\mu+1)} e^{-\beta(E_\nu-E_\mu)}, &
\frac{p_\nu}{p_\mu}<1\\
1,  & {\rm  otherwise.}
\end{array}\right.
\end{eqnarray}
The MCBA algorithm is complete and the three
basic steps are repeated a number of times. After an initial
equilibration time, the expectation values can be then estimated
as an arithmetic mean over the repeated Markov chain
\begin{eqnarray}
\langle Q\rangle=\frac{1}{N}\sum_{\{\mu\}}Q(\mu).
\end{eqnarray}

\begin{figure}
\includegraphics{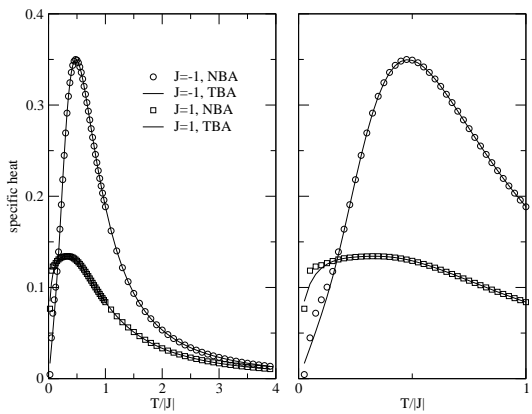}
\caption{The specific heat of 24 sites
anti-ferromagnetic and ferromagnetic XXX model (points) and the
same quantities obtained by TBA (lines).} \label{mcba_FIGURE_CV24}
\end{figure}

\begin{figure}
\includegraphics{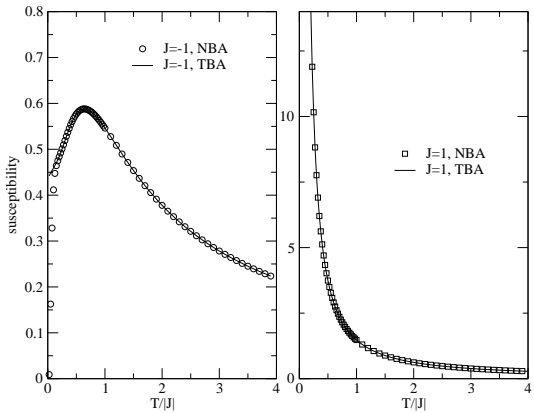}
\caption{The susceptibility of 24 sites anti-ferromagnetic (left)
and ferromagnetic (right) XXX model (points) and the same
quantities obtained by TBA (lines).} \label{mcba_FIGURE_SUS24}
\end{figure}

\begin{figure}
\includegraphics{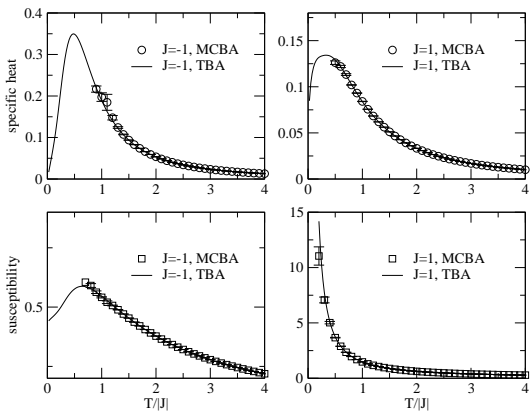}
\caption{The specific heat and susceptibility for  a 60 sites
chain, computed with the MCBA method, is compared with the TBA
results ({\bf lines}) both for the anti-ferromagnetic ({\bf
circles}) and ferromagnetic ({\bf squares}) cases.}
\label{mcba_FIGURE_MCBA}
\end{figure}

\section{Specific Heat and Susceptibility}
\label{mcba_sec:spce}

In order to check the validity of
our approaches, we apply these two methods to
the study of the specific heat and of the magnetic susceptibility of
the
anti-ferromagnetic and ferromagnetic Heisenberg models.

For the present model, however, because of the degeneracy in each
set of quantum number configuration, Eq.
(\ref{mcba_eq:expectationQ}) should be revised according to the
property of the operator. For example, the internal energy and
magnetization are
\begin{eqnarray}
\langle E\rangle=\frac{1}{Z} \sum_{\mu}(N-2M_\mu+1)E_\mu e^{-\beta
E_\mu},
\nonumber \\
\langle M\rangle=\frac{1}{Z}
\sum_{\mu}\sum_{M_\mu^z=-N/2+M_\mu}^{N/2-M_\mu}2M_\mu^z e^{-\beta
E_\mu}. \label{mcba_eq:UMexpectation}
\end{eqnarray}
where $Z=\sum_{\mu}(N-2M_\mu+1) e^{-\beta E_\mu}$. From
thermodynamics it is easy to have the expression for specific heat
and magnetic susceptibility per site
\begin{eqnarray}
C=\frac{\beta^2}{N}(\langle E^2\rangle - \langle E
\rangle^2),\nonumber \\
\chi=\frac{\beta}{N}(\langle M^2\rangle - \langle M \rangle^2)
\label{mcba_eq:CVSUS}
\end{eqnarray}

\begin{table}
\label{mcba_table2} \caption{Specific heat and susceptibility of
ferromagnetic XXX model obtained by thermodynamic Bethe ansatz
(TBA), numerical Bethe ansatz (NBA) solution of 24 sites, and
Bethe ansatz based Monte Carlo (MCBA) approach for 60 sites
system.}
\begin{center}
\begin{ruledtabular}
\begin{tabular}{c|lccl}
 &$T/J$ & TBA & NBA & MCBA
\\\hline
$C_v$& 0.5 & 0.129178  & 0.129197  &  0.1261 $\pm$ 0.0019   \\
& 0.6 & 0.121671  & 0.121688  &  0.1220 $\pm$ 0.0012   \\
&0.7 & 0.112363  & 0.112379  &  0.1132 $\pm$ 0.0009   \\
&0.8 & 0.102496  & 0.102511  &  0.1019 $\pm$ 0.0003    \\
&0.9 & 0.092861  & 0.092875  &  0.0931 $\pm$ 0.0002   \\
&1.0 & 0.083875  & 0.083887  &  0.0840 $\pm$ 0.00017  \\
&1.5 & 0.051111  & 0.051117  &  0.05112 $\pm$ 0.00004   \\
&2.0 & 0.033256  & 0.033260  &  0.03326 $\pm$ 0.00002   \\
&2.5 & 0.023088  & 0.023090  &  0.02308 $\pm$ 0.00001  \\
&3.0 & 0.016876  & 0.016878  &  0.01687   \\
\hline $\chi$ &0.5 & 3.7378    & 3.742446  & 3.669 $\pm$ 0.027  \\
& 0.6 & 2.90686   & 2.909917  & 2.8914 $\pm$ 0.0139    \\
&0.7 & 2.35856   & 2.360709  & 2.3458 $\pm$ 0.0039   \\
&0.8 & 1.97323   & 1.974817  & 1.9613 $\pm$ 0.0044  \\
&0.9 & 1.68953   & 1.690744  & 1.6902 $\pm$ 0.0028   \\
&1.0 & 1.473032  & 1.473986  & 1.47509 $\pm$ 0.00180   \\
&1.5 &  0.882613  & 0.882996  & 0.88316 $\pm$ 0.00016   \\
&2.0 &  0.622855  & 0.623056  & 0.62299 $\pm$ 0.00010  \\
&2.5 &  0.479082  & 0.479205  & 0.47908 $\pm$ 0.00007 \\
&3.0 &  0.388433  & 0.388516  & 0.38848 $\pm$ 0.00004   \\
\end{tabular}
\end{ruledtabular}
\end{center}
\end{table}

\begin{figure*}

\includegraphics{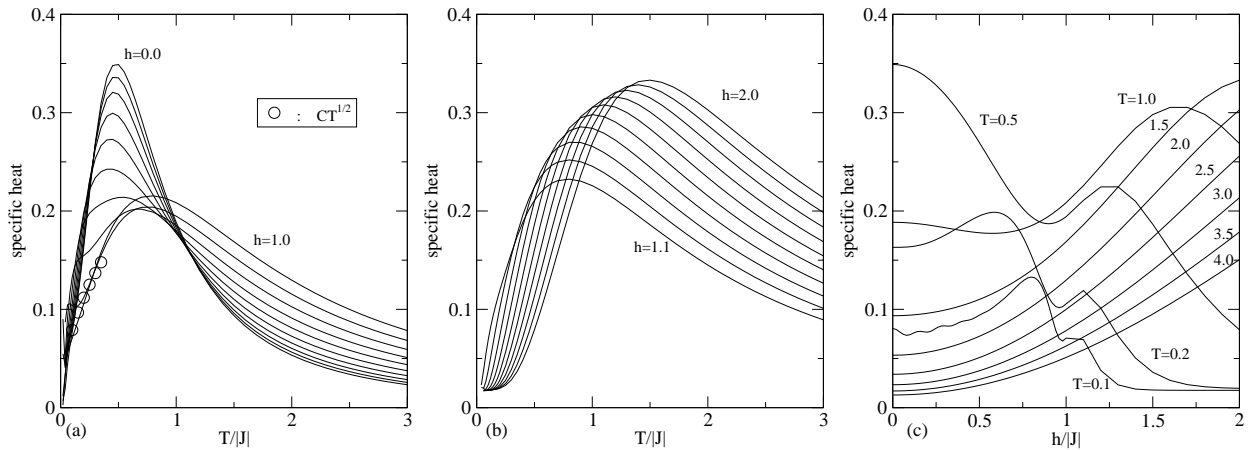}
\caption{\label{fig:acvh} The specific heat of the
anti-ferromagnetic Heisenberg model for different values of the
external field: (a) $h=0, 0.2, 0.3, \dots, 1.0$; (b) $h=1.1, 1.2,
\dots, 2.0$; (c) specific heat as a function of $h$ for different
temperature $T=0.1, 0.2, \dots, 4.0$.
In panel (a) of fit to the law $C=\propto T^{1/2}$, for h=1.0, is given
at low temperatures.
\\\\}
\end{figure*}

\begin{figure*}

\includegraphics{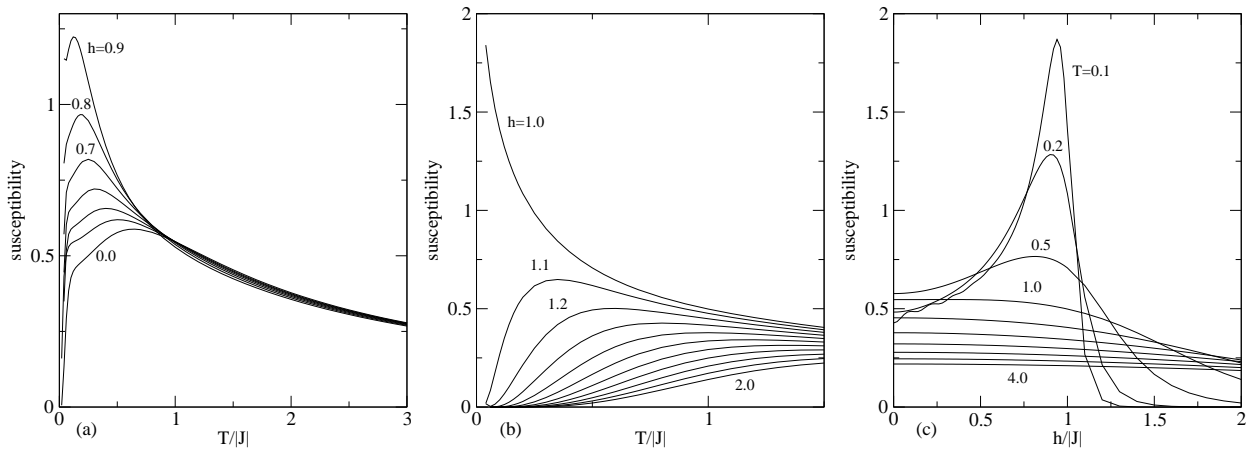}
\caption{\label{fig:asuh} The magnetic susceptibility of the
anti-ferromagnetic Heisenberg model for different values of the
external field: (a) $h=0, 0.4, 0.5, \dots, 0.9$; (b) $h=1.0, 1.2,
\dots, 2.0$; (c) magnetic susceptibility as a function of $h$ for
different temperature $T=0.1, 0.2, \dots, 4.0$.}
\end{figure*}

We apply NBA to 24-site system and MCBA to 60-site system,
respectively. The later has $C^{60}_{30}$ different quantum number
configurations, hence it is impossible to calculate all the eigenvalues
of the system.

In Figs. \ref{mcba_FIGURE_CV24} and \ref{mcba_FIGURE_SUS24}, we
show the specific heat and the   magnetic susceptibility, for a
24-site system, obtained from NBA and compare our results with those
obtained from TBA. It is clear that the two results match. In Fig
\ref{mcba_FIGURE_MCBA}, we show the specific heat and the magnetic
susceptibility, for a  60-site system obtained from MCBA together
with the results from TBA. They both agree to each other except
at low temperature. In Table \ref{mcba_table2}, we compare, for
the ferromagnetic case, the two methods we introduced here with
TBA, giving the explicit numerical values. It is clear that our
methods work very well for the present model. Hence our conclusion
is that for  small systems, such as $N\leq 38$, and for the
Heisenberg chain, it is possible to compute all eigenvalues
 and  to obtain all possible thermodynamic
quantities of interest by using Eq. (\ref{mcba_eq:expectationQ}).
For temperatures larger than the finite size energy gap our results
agree with TBA results exactly. For larger systems, however, results
can still be obtained by using the MCBA method.

\begin{figure*}
\includegraphics{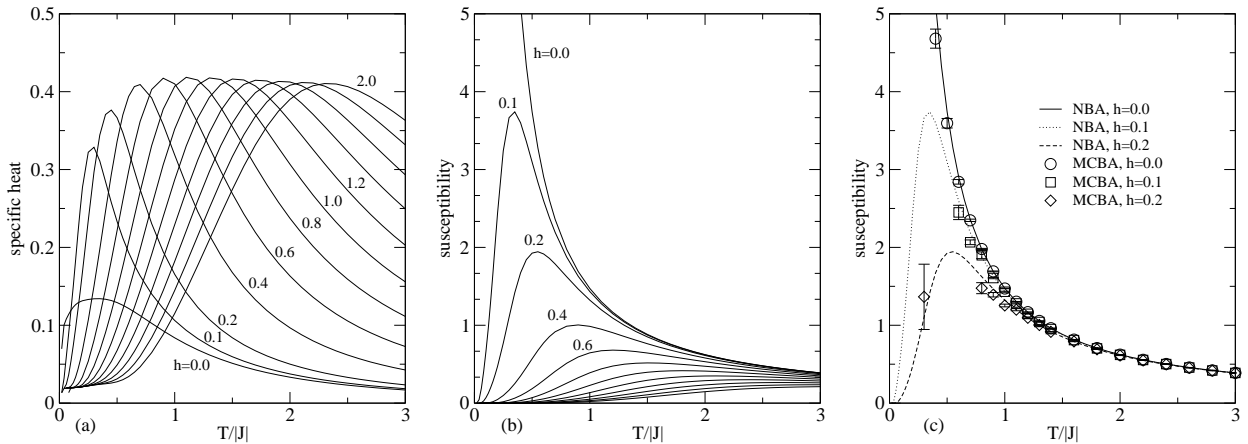}
\caption{\label{fig:feffo}The specific heat(a)  and the magnetic
susceptibility(b)  ferromagnetic Heisenberg model as a function of
temperature for different values of external field $h=0.0, 0.1,
0.2, \dots, 2.0$, obtained by NBA. (c). The susceptibility of
ferromagnetic Heisenberg model for $h=0.0, 0.1, 0.2$, obtain by
NBA and MCBA respectively.}
\end{figure*}

Now we study the thermodynamics of the model in the presence of
a magnetic field by NBA, which has also been studied by
Kl\"umper \cite{AKlumper98}. In Fig. \ref{fig:acvh} and Fig.
\ref{fig:asuh}, the results for the specific heat and the magnetic
susceptibility of the anti-ferromagnetic case are shown for various
magnetic fields. It is clear from these two figures that there are
two different behaviors at low temperature, separated by the
saturation field $h_c=1.0$ at the ground state. In order to understand
better this behavior of the antiferromagnetic case, let us use the mapping
between the Heisenberg model and the spinless fermion model.
This mapping is achieved by
the Jordan-Wigner transformation \cite{EFradkinb},
and Hamiltonian (\ref{mcba_eq:Hamitonian}) can be written as
\begin{eqnarray}
{\cal H} &=& -\frac{J}{2}\sum_{l=1}^N(f^\dagger_l
f_{l+1}+f^\dagger_{l+1} f_l) \nonumber \\ &&
+J\sum_{l=1}^N(n_l-\frac{1}{2})(n_{l+1}-\frac{1}{2}).
\end{eqnarray}
where the spinless fermion operators $f^\dagger_l, f_l$ obey the
usual anti-commutation relation, $n_l$ is the usual local number
operator. When $h<h_c$, the system is not fully polarized, that is
$\sum_{l=1}^N n_l>0$, hence we always have two Fermi points $\pm
k_F$ at the ground state. The dispersion relation of low-lying
excitations is dominated by the linear-$k$ dependence, hence we
still have the Fermi-liquid like specific heat: $C\propto T$ at
low temperatures. If $h\geq h_c$, however, and from the point view
of spinless fermions, we have $\sum_{l=1}^N n_l=0$, and  the
dispersion relation becomes $k^2$, because of the $\cos k$
dispersion-relation for the fermions in the lattice. Hence, the
specific heat manifest a $T^{1/2}$ behavior at sufficiently low
temperature for $h=h_c$, which can be seen in Fig. \ref{fig:acvh}
(open circles). Moreover, the magnetic susceptibility presents a
strong peak for $h=h_c$, when $T\rightarrow 0$
[see Fig. \ref{fig:asuh}, panel (c)]. This strong magnetic
response is associated with a change in the nature of the elementary
excitations when the line $h=h_c$ is crossed at zero temperature.
Indeed, at $T=0$ and  $h_c=1.0$,  the system manifest infinite
susceptibility, as can be seen from Fig. \ref{fig:asuh},
panel (b). We attribute it due to the degeneracy between the
state of $[N-1,1]$ and $[N]$, and a small magnetic field can fully
polarize the system. The phase with $h\ge h_c$ share
anti-ferromagnetic-like behavior [Fig. \ref{fig:feffo}, panel (b)], while
for $h<h_c$, the susceptibility shows a logarithm
singularity \cite{AKlumper98}.

For the ferromagnetic case
 the specific heat and the magnetic susceptibility are ploted 
in Fig. \ref{fig:feffo}, for different values of the magnetic field. As
is known, if $h=0$ the ground state of the ferromagnetic case is
highly degenerate with $S=N/2, S_z=-S, -S+1, \cdots, S$ and a very
small $h$ can fully polarized the system. So it is easy to
understand why zero temperature susceptibility is infinite. After
it is magnetized (in the presence of small $h$), however, the
susceptibility should be zero. This behavior is seen in Fig.
\ref{fig:feffo}, panel (b). We also show, in Fig.
\ref{fig:feffo}, panel (c), the susceptibility obtained by MCBA. Both the
results of the two methods  agree with each other perfectly.

\section{Summary}
\label{mcba_sec:summary}

In summary, we presented two numerical approaches to discuss the
thermodynamics of Bethe ansatz solvable models. The
first one is the numerical Bethe ansatz which works very well for a
small system. We think it is possible to obtain all eigenvalues of
a system up to size $L=38$, for the Heisenberg model. 
For a relatively larger system,
we also find that the Monte Carlo simulation in quasi-momentum space
works well in the moderate
and high temperature regions. At low temperatures
the present selection method
is not excellent, and a better one is required. The discovery 
of such a method is a challenging and interesting research problem.

There are many physical quantities of interest at finite
temperature which are still not well understood, such as spin
stiffness of XXZ model, important to understand the transport
properties, because of the complex form of the thermodynamic
equations. Our methods provide a new route to compute all  these
quantities directly from the Bethe ansatz equations.

This work is supported by trans-century projects and Cheung Kong
projects of China Education Ministry and
 by the Portuguese Program
PRAXIS XXI under grant number 2/2.1/FIS/302/94. We thank M.
Takahashi for sending us the TBA data. SJG and NMRP want to thank
the support of the Physics Department of the University of
\'Evora, where part of this research was done.
 SJG would like to thank
Dr. M. B. Luo for helpful discussion on Monte Carlo methods.

\end{document}